%% file: main.tex
\documentclass[lettersize,journal]{IEEEtran}
\IEEEoverridecommandlockouts
\usepackage{cite}
\usepackage{amsmath,amssymb,amsfonts}
\usepackage{algorithmic}
\usepackage{algorithm}
\usepackage{array}
\usepackage{graphicx}
\usepackage{textcomp}
\usepackage{xcolor}
\usepackage[caption=false,font=normalsize,labelfont=sf,textfont=sf]{subfig}
\usepackage{stfloats}
\usepackage{url}
\usepackage{cite}
\usepackage{tikzsymbols}
\usepackage{tikz}
\tikzset{>=latex}
\usetikzlibrary{
    external,
    patterns,
    patterns.meta,
    arrows,
    positioning,
    chains,
}
\usepackage{pgfplots}
\usepgfplotslibrary{groupplots}
\usepgfplotslibrary{dateplot}
\pgfplotsset{compat=1.18}
\usepackage{microtype}
\usepackage{subcaption}
\usepackage{multirow}

\usepackage{tabularx}
\usepackage{xcolor}

\usepackage{glossaries}
\loadglsentries{acronyms}

\begin{document}

\title{Quantum Backbone Networks for Hybrid Quantum Dataframe Transmission}

\author{
    \IEEEauthorblockA{Francesco Vista},   
    \IEEEauthorblockN{Daniel Holme}, and
    \IEEEauthorblockN{Stephen DiAdamo}
    \thanks{Francesco Vista and Stephen DiAdamo are with the Cisco Quantum Lab in Munich, Germany. Francesco Vista is also with Politecnico di Bari. Daniel Holme is with Cisco Systems.
    
    \textcopyright \hspace{1mm} 2024 IEEE. Personal use of this material is permitted. Permission from
    IEEE must be obtained for all other uses, in any current or future media,
    including reprinting/republishing this material for advertising or promotional
    purposes, creating new collective works, for resale or redistribution to servers
    or lists, or reuse of any copyrighted component of this work in other works.}
}

\maketitle

\begin{abstract}
    To realize a global quantum Internet, there is a need for communication between quantum subnetworks. To accomplish this task, there have been multiple design proposals for a quantum backbone network and quantum subnetworks. In this work, we elaborate on the design that uses entanglement and quantum teleportation to build the quantum backbone between packetized quantum networks. We design a network interface to interconnect packetized quantum networks with entanglement-based quantum backbone networks and, moreover, design a scheme to accomplish data transmission over this hybrid quantum network model. We analyze the use of various implementations of the backbone network, focusing our study on backbone networks that use satellite links to continuously distribute entanglement resources. For feasibility,  we analyze various system parameters via simulation to benchmark the performance of the overall network. 
\end{abstract}


\section{Introduction} \label{sec:intro}
\input{Sections/01_intro}

 \section{Quantum Backbone Networks} \label{sec:backbone}
\input{Sections/02_backbone}

\section{Satellite Quantum Backbone Networks} \label{sec:satellite-comm}
\input{Sections/03_satellite_comm}

\section{Quantum Network Ingress and Egress Design} \label{sec:interface}
\input{Sections/04_interface}

\section{Simulation Results} \label{sec:results}
\input{Sections/05_results}

\section{Implementation Challenges} \label{sec:challenges} 
\input{Sections/06_challenges}

\section{Conclusion} \label{sec:conclusions}
\input{Sections/07_conclusion}

\bibliographystyle{IEEEtran}
\small{\input{ref.bbl}}

\input{bios}

\end{document}

%% file: acronyms.tex
\newacronym{TLE}{TLE}{Two-Line Element}
\newacronym{LEO}{LEO}{Low Earth Orbit}
\newacronym{MEO}{MEO}{Middle Earth Orbit}
\newacronym{GEO}{GEO}{Geostationary Earth Orbit}
\newacronym{HAPS}{HAPS}{High-Altitude Pseudo-Satellite}
\newacronym{FSOL}{FSOL}{Free-Space Optical Links}
\newacronym{QKD}{QKD}{Quantum Key Distribution}
\newacronym{ISL}{ISL}{Inter-Satellite Links}
\newacronym{MPLS}{MPLS}{Multi-Protocol Label Switching}
\newacronym{OS}{OS}{Optical Switch}
\newacronym{qubit}{qubit}{quantum bit}
\newacronym{UAV}{UAV}{Unmanned Aerial Vehicle}

%% file: Sections/01_intro.tex
A global quantum network can enable advancement in secure communication and quantum computing capabilities. Different types of quantum network designs are available, such as \gls{QKD}, entangled-based with teleportation, and packetized quantum networks, for some. The first kind are employed only to distribute secret keys between users, such that the security of the key is theoretically provable. The second, instead, is a network where no quantum information travels through the quantum channels. Specifically, at the core of this networks is the distribution of entanglement, a key resources to perform the quantum teleportation protocol. Entanglement is generated by network resources and distributed to quantum nodes through fiber or free space communication channels. To transmit the quantum state information is done by consuming the entanglement resource and communicating classical information afterwards. 

Although it is possible to distributed quantum entanglement and transmit quantum states over short distance, there is a limitations in reachable distance. This implies that the deployment of a large-scale quantum network requires an analog of a repeater, known as a quantum repeater. A quantum repeater is used to extend the range of quantum transmission. Quantum repeaters are responsible for creating entanglement links between adjacent nodes and performing entanglement swapping to establish end-to-end entanglement and use quantum teleportation to transmit quantum information classically. A drawback of using quantum repeaters is the communication rate does not scale with the number of users~\cite{coutinho2022robustness}. On the other hand, scale-free networks based on packetized quantum communication~\cite{munro22designing, diadamo2022packet, yoo2021wrapper}, the third type of proposed quantum network, could support more users \cite{Mandil2023} but at the cost of reachable network radius. What has been proposed, therefore, is the concept of connecting quantum subnetworks based on quantum packets with a quantum backbone network \cite{de2023satellite}. What results is a network---the backbone---to distribute quantum entanglement over large distances, but with scale-free networks at the subnetwork level, using the backbone to perform inter-network quantum communication. 

Since single-photon transmission over fiber suffers high loss, increasing exponentially with distance, to extend to long range, many quantum repeaters are needed to scale large distances. A viable near-term solution to overcome this is to use satellite communication. A pivotal milestone  for quantum satellite communication was the launch of the Micius satellite, which showcased the feasibility of quantum key distribution and quantum teleportation protocols \cite{lu_micius_2022}. In fact, since free-space transmission follows a square-loss law, the use of satellite constellations can be a better solution for interconnecting distant quantum networks, especially in the near term. For this reason, the scientific community has investigated the performance in terms of quantum key distribution rates considering a satellite as backbone network \cite{yehia2023connecting} and in terms of entanglement end-to-end rate considering a constellation made up with more than one satellite \cite{chiti2021quantum}. To our knowledge, no other work proposes a design of a network interface for merging packetized quantum networks or an analysis of the architecture for transmission of \glspl{qubit} between subnetworks using hybrid methods of satellite and fiber.

Building upon these considerations, we present a quantum network backbone design that seamlessly integrates satellite and fiber links, thereby ensuring a continuous and robust entanglement service. In this regard, this work provides the following contributions: 1) A design of a quantum backbone network for hybrid quantum dataframe transmission; 2) A network interface design to merge a packetized quantum network and an entanglement-based backbone network; 3) A performance analysis using different \gls{LEO} satellites and a direct fiber links between remote locations as quantum backbone network.

In particular, the performance of our design and protocol is evaluated in terms of the number of qubits received---the coincidence count---under different quantum memory sizes. We model two packetized quantum networks connected by a quantum backbone network. What we find is, using a combination approach for generating entanglement with the backbone to the subnetwork edges, alternating the use of the fiber and satellite network when the satellites are visible, can lead to an enhanced quantum communication rate between subnetworks.


%% file: Sections/02_backbone.tex
A backbone network is a network used for interconnecting subnetworks. It is usually responsible for long-haul network traffic and has high-capacity channels to transmit with high transmission rates. Backbone networks generally use different routing schemes and protocols than the subnetworks to maximize throughput between subnetworks. Within the subnetworks, there are ingress and egress nodes at the network edges responsible for interfacing with the backbone network. In quantum networking, the applications of a quantum backbone network are similar. A quantum backbone network is used to interconnect multiple quantum subnetworks and features corresponding ingress and egress nodes for interfacing.

In recent work \cite{diadamo2022packet, munro22designing}, the concept of a quantum backbone network with hybrid routing has been introduced. Packet-switched quantum networks use hybrid classical-quantum dataframes with direct transmission of quantum states~\cite{diadamo2022packet}. The hybrid frames pre- and post-pend a quantum payload with classical routing and error correction information. This information is used to dynamically switch the quantum payload through the network. On the other hand, an entanglement-based quantum network relies on stored entanglement, consumed to perform quantum teleportation, to transmit quantum information. Quantum teleportation is a quantum communication protocol that can transmit quantum information indirectly by consuming one maximally entangled pair of qubits and classical communication. This allows important quantum information to be reliably transmitted, as it is never directly sent over a lossy channel, rather only classical information about the state is sent.

At the metropolitan scale, using packet-switched quantum networks offers advantages by eliminating the need for entanglement distribution and robust quantum memories, thereby obviating the use of teleportation. However, a significant challenge arises: without a quantum repeater, distance limitations become unavoidable, and beyond roughly 100 km of standard fiber, the quantum communication rate approaches zero. On the contrary, entanglement-based networks aim to address this issue, albeit with drawbacks. While they resolve distance limitations, the communication rate is significantly reduced compared to direct transmission at short distances, bounded by the rate of entanglement generation. Additionally, for multiple hops, the protocol requires high synchronization among nodes for entanglement swapping, impacting the scalability of users the network can support~\cite{coutinho2022robustness}.

\begin{figure}
    \centering    
    \includegraphics[]{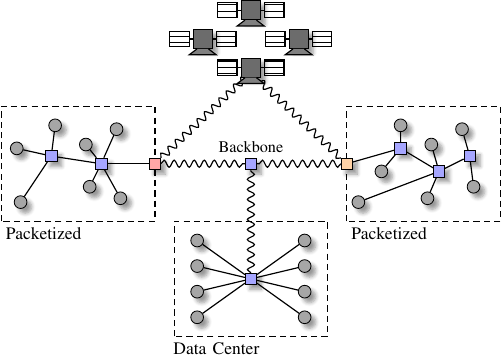}
    \caption{\small{Two packetized networks and a data center are interconnected via an entanglement-based backbone network. The network backbone can be composed of fiber technology or satellite.}}
    \label{fig:backbone-network}
\end{figure}

To overcome this problem, proposed in \cite{diadamo2022packet, munro22designing} is to merge the two network types into a single network, but an explicit design and protocol were not explored. In this work, we use the general strategy of using packet switching at the access level of the network, or the subnetworks, making use of dynamic switching with as large a network radius as possible. At the maximum transmission subnetwork radius, an egress node is placed as one end of a backbone network. The backbone network’s objective is to have entanglement already established between the egress and ingress of the destination network such that when quantum information arrives at the egress, the quantum information can be teleported immediately. In Fig.~\ref{fig:backbone-network} we depict the setting, where multiple subnetworks share the same backbone network.

The backbone network can use various technologies to perform long-haul communication between subnetworks.  A backbone network can be composed of fibre connections, a terrestrial network, it can be composed of satellite links, using the edge nodes as ground stations, forming a non-terrestrial network, or it can be a combination of both. This backbone network's job is to generate entanglement at the entry points of the subnetworks so it is available when teleportation is needed. To achieve that, the backbone network uses the techniques that have been studied for entanglement distribution, swapping, and purification to achieve end-to-end entanglement. In this work we explore the trade-offs between the various link-types. Specially, as a starting point to creating a backbone link, we analyze a single node backbone network that connects to the subnetworks via fiber links, \gls{LEO}, \gls{MEO}, or \gls{GEO} satellite. 

Indeed, the subnetwork structure need not be restricted to packet-switched networks. The backbone network can connect any type of quantum subnetwork, be it a quantum datacenter, entanglement-based, or another kind. The two essential components to integrate another network type would be to design the backbone network interface to function according to the subnetwork and define the protocols for de- and re-constructing the dataframes. This enables heterogeneous networks where subnetworks based on a different types of quantum networks can communicate using their own routing protocols locally, using the backbone network to communicate with other networks.

%% file: Sections/03_satellite_comm.tex
Given notable advances in satellite technology and declining launch costs, deploying satellites for communications is becoming commonplace. With various available orbits, recent quantum entanglement sources have been launched into \gls{LEO}, \gls{MEO}, and soon \gls{GEO}. 
Despite \gls{MEO} and \gls{GEO} satellites offering extended service windows, they are too distant from ground stations, leading to potential high channel loss \cite{de2023satellite}. For nearer-term considerations, our analysis focuses solely on \gls{LEO} satellites, reviewing quantum communication in \gls{LEO} satellite networks and exploring quantum network formation via a \gls{LEO} satellite constellation.

\subsection{Single LEO Satellite Backbone}

Numerous simulations and real-world experiments confirm the effectiveness of entanglement distribution via satellite links in \gls{LEO}, approximately 500 to 1,000 km~\cite{lu_micius_2022}. For satellite entanglement distribution, an onboard entanglement source emits entangled photons sent through Free-Space Optical Links (FSOL) using telescopes. The entanglement source can be spontaneous parametric down conversion or quantum dot-based technology. In the downlink scenario, atmospheric effects on the entangled photon's journey to Earth are predominantly encountered at the end, following near-vacuum space travel where the beam maintains diffraction limit properties \cite{de2023satellite}. Notably, the optical beam-width of the ground station is significantly distorted when initially passing through the turbulent environment.

The downlink optical beam faces atmospheric turbulence in its final path segment. As it enters the atmosphere, the satellite's beam-width surpasses the turbulent vortex size, minimizing its impact on propagation due to the standard aperture radius. Atmospheric effects on signal propagation hinge on sky conditions and transmission wavelength, assessable through simulators like Lowtran (utilized in \cite{yehia2023connecting}). When the satellite nears the horizon, the optical refractive index variation causes beam deviation, elongating the optical path—significant only for elevation angles below 20 degrees \cite{de2023satellite}. In the downlink scenario, random variations predominantly result from pointing and tracking errors due to the rapid distance changes between LEO satellite constellations and ground stations (traveling at 7.8 km/s). Mitigating errors requires high-precision alignment technology with closed-loop motion control.

Ensuring synchronization between two ground stations receiving entangled photons from the satellite is crucial for quantum teleportation. Due to different latencies, only matching entangled pairs are viable. To address latency variations, a dynamic delay, facilitated by a varying fiber delay line or a quantum memory, is required for one of the photons. The calculation of distance between the satellite and ground stations, considering ephemeral numbers adjusted for velocity, trajectory, and time sequence, determines the necessary delay time. A methodology for continuous computation of delay, applicable on either the satellite or the ground station, is essential for maintaining entanglement synchronization between the edge nodes.

For inter-network quantum communication, entangled photons must be stored in destination networks for later consumption through a quantum teleportation process. Achieving this requires highly accurate time synchronization between egress and ingress nodes. In the teleportation process, sibling photons received from the satellite are consumed on both sides, necessitating precise clock synchronization for the parties to match pairs. Current technology offers well-known protocols capable of clock synchronization with low nanosecond accuracy, extending to picoseconds. These protocols can be directly employed. Given that teleportation relies on perfectly synchronized quantum memories in the egress and ingress, the critical factor is the accuracy of the clocks.

A single satellite facilitating entanglement distribution simplifies the process, but quantum communication in space, albeit less constrained than fiber communication, still faces distance limitations. Building a global quantum internet will thus demand the coordinated efforts of multiple satellites to establish an extensive quantum backbone network. Despite the numerous considerations involved, we offer a conceptual outline of the prerequisites for such a network in the next subsection.

\subsection{Multiple LEO Satellite Backbone}

In the context of a single \gls{LEO} satellite, addressing physical constraints, the constellation resource manager utilizes Inter-Satellite Links (ISL) to establish a route between edge nodes, even when the remote node is beyond the horizon and out of sight. This resembles the use of a quantum relay or repeater in terrestrial networks. The satellite must execute either a pure optical switching path to prevent photon measurement or employ a quantum repeater protocol for maintaining the quantum state over a long distance. The quantum repeater protocol can involve entanglement swapping or a quantum error correction technique to establish multi-hop entanglement within the satellite constellation~\cite{muralidharan2016optimal}.

Classically, in LEO constellations ``path routing" is typically decided and executed from a Software Defined Networking (SDN) controller utilizing source-based routing across the network using labels in \gls{MPLS} Segment Routing packets, or similar, to identify and select each hop in the path through space and then insert the path in the label stack to be executed as the packet traverses the network. This technique is well tested both in terrestrial networks and space. In order to create the short-lived routes in space for a short time for entanglement swapping (changes are required roughly every 7-10 minutes in LEO) a similar capability is required.


Once a path is established, a similar protocol to the quantum repeater can be used. In \cite{boone2015}, a first attempt has been made to overcome the distance limitation by proposing a hybrid space-ground quantum repeater scheme. In particular, they envision a network composed of a satellite equipped with an entanglement source, and ground repeaters with quantum non-demolition (QND) measurement devices and quantum memories. Here, the satellite continuously generates and transmits entangled photons to both ground quantum repeaters. Once the photons are successfully received, they are stored in the quantum memory, and entanglement swapping executes to extend the end-to-end entanglement. This scheme is not scalable since it requires that each ground station has good weather conditions, which is often not the case. Recently, a fully space-based quantum repeater scheme has been proposed \cite{gundougan2021}. In this case, satellites are also equipped with QND devices and quantum memories to perform entanglement swapping in space. This approach offers the advantage of having inter-satellite communications with low channel loss.

%% file: Sections/04_interface.tex
\begin{figure}
    \centering        
    \includegraphics[]{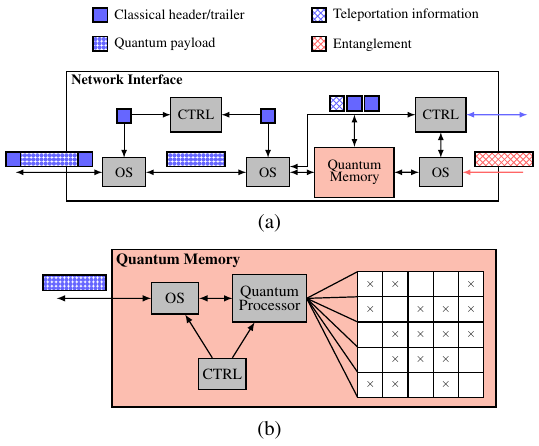}
    \caption{\small{A hybrid quantum network interface design. In (a) is a depiction of the processing of a hybrid frame. The quantum payload is processed in the memory depicted in (b). The memory stores individual entangled qubits marked by $\times$, where some of them may be lost in transmission.}}
    \label{fig:network-interface}
\end{figure}

To support the communication schemes between quantum subnetworks through the quantum backbone network, quantum ingress and egress nodes will be needed. The nodes act to facilitate quantum communication based on different protocols. In this work, we focus on the interface between a packetized quantum network~\cite{diadamo2022packet} and an entanglement-based backbone network. To facilitate the merger of the two, we propose a network interface design in Fig.~\ref{fig:network-interface}.  

Transmission between subnetworks operates by employing packet-switching or, in the nearer-term, a burst-switching approach, as detailed in~\cite{diadamo2022packet, Mandil2023}. Hybrid classical-quantum frames are forwarded to a network egress, utilizing classical header information for routing. The egress serves as the interface to the backbone network, equipped with pre-established entanglement resources to the desired subnetwork. The hybrid frame undergoes header and trailer splitting, and the quantum payload is teleported through the channel. Classical teleportation information, along with header and trailer details, is transmitted via classical means using circuit switching over the backbone network to minimize latency. At the destination subnetwork, an ingress node receives classical state information, reconstructs qubits locally, and retrieves header and trailer details. The ingress node reframes the quantum payload,  transmitting header and arriving qubits, then utilizes packet-switching to forward the frame to its destination.

In Fig.~\ref{fig:network-interface}(a), the ingress and egress are depicted. In the upper portion illustrating egress behavior, an incoming hybrid classical-quantum frame arrives from the left. An \gls{OS} splits the payload from the header and trailer. The header and trailer are processed by a classical controller to determine the payload's destination, and further control information is sent forward to be sent outward. In this step, burst switching can also used to add additional time for the OSs and the payload may arrive behind the header with a time delay, which can reduce the loss of the payload~\cite{Mandil2023}. The payload, instead, is routed to the quantum memory for the teleportation process. The header, trailer, and teleportation information, all classical data, are then sent through the backbone network, exiting from the right side of the figure.

To perform as a network ingress, an incoming signal would arrive from the right side of the figure and the process would run in reverse. The classical messages pass through a controller, where the header and trailer go forwarding in the frame reconstruction phase and the teleportation data is fed into the quantum memory. The teleportation information is used to complete the teleportation protocol, where the quantum information is reconstructed and emitted onto an output fiber. When that is complete, the header and trailer are framed around the output quantum payload and sent onward into the subnetwork.

The quantum memory, depicted in Fig.~\ref{fig:network-interface}(b), is a complex device comprising various components. When a quantum payload enters the quantum memory, it traverses an optical switch to guide it to the correct part of a quantum processor to execute the quantum teleportation protocol. To perform quantum teleportation, pre-existing shared entanglement resources must be stored and available when the protocol starts. The quantum memory storage must be indexed, as entanglement resources alone lack identifying information. Furthermore, the storage index must synchronize with a corresponding index at the ingress of the destination subnetwork, ensuring the correction bits for teleportation are applied to the corresponding parts of the entangled pairs. Various queuing practices can achieve this synchronization~\cite{diadamo2023impact}. During the teleportation protocol, classical bits are generated, and payload qubits and entanglement resources are consumed through quantum measurement, freeing up a memory slot in the storage unit. The classical bits, along with information on storage unit indexes, exit the memory to the interface controller to be forwarded onward.

On the incoming side, classical information can enter the quantum memory in order to reconstruct the quantum payload. This information is the classical teleportation information that is output from the above process. Here, the correction bits from the teleportation protocol are applied to the synchronized index using the quantum processor. Once applied, the qubits are released from memory and sent out from the memory through an \gls{OS}.

In addition to being used to perform teleportation, the quantum memories in the edge nodes of different subnetworks always accept entangled qubits between themselves using the quantum backbone network. Entanglement units (along with some classical information) enter the quantum memory for storage. The classical information can be used to identify with who the other half of the entangled pair is stored. A protocol for entanglement distribution at the network level would be used here to synchronize the nodes and build up entanglement resources.

%% file: Sections/05_results.tex
In this Section, we study the feasibility of our quantum backbone network design and protocol by investigating the achievable performance in terms of the successful number of qubits exchanged between two nodes belonging to different remote subnetworks. In particular, we consider two subnetworks in which the egress node, depicted as the red square in Fig.~\ref{fig:backbone-network}, is situated in Munich ($48^\circ 09'$ N, $11^\circ 32'$ E), while the ingress node, the orange square in Fig.~\ref{fig:backbone-network}, is in Nuremberg ($49^\circ 26'$ N, $11^\circ 07'$ E). The distance between egress and ingress nodes is approximately 150 km.

To define our quantum backbone network we use two configurations. The first consider a single \gls{LEO} satellite, while the second a ground source placed equidistantly between the egress and ingress nodes of the subnetworks.
For the satellite backbone, we explore three distinct LEO satellite orbits, namely 1) Micius, 2) Iridium-126, and 3) Starlink-2007 orbits, as outlined in Table~\ref{tab:sats}. To gather relevant data for each satellite, we discretize the orbit into 2-second intervals, varying the parameters at each step. For every step, we collect essential information, including the elevation angle, channel length, and the channel attenuation. This information is obtained using the satellite models defined in~\cite{yehia2023connecting}. For the ground-based backbone, we consider the impact of standard optical fiber and dark fiber with attenuation coefficient of $0.2$ dB/km and $0.16$ dB/km, respectively.

We model the network and its behavior using discrete event simulation as follows. A channel of 5 km standard fiber with channel loss is established between a quantum node, within the first subnetwork, and the Munich egress for simulating network traffic. Hybrid frames are generated by the quantum node at a rate of $100$ MHz and a frame duration of $100$ $\mu s$, producing 100,000 qubits per frame. Frames are transmitted over the channel following a Poisson process with an average inter-arrival time of $20$ ms. The quantum backbone network, instead, continuously generates entanglement for the egress/ingress nodes, emitting pairs at a rate of 0.2 MHz. Entanglement loss occurs based on channel attenuation. Successfully arriving at ingress and egress nodes, entanglement is stored in a quantum memory of $M$ slots (see Fig.~\ref{fig:network-interface}). Upon the frame reaching the Munich egress, each qubit is teleported to the Nuremberg ingress with a 50\% success rate, and classical information is sent via direct fiber connection to the ingress. Subsequently, the reconstructed hybrid frame is forwarded to the destination node within the secondary subnetwork, travelling 5 km of standard fiber channel with loss.

\begin{figure}
    \centering
    \includegraphics[]{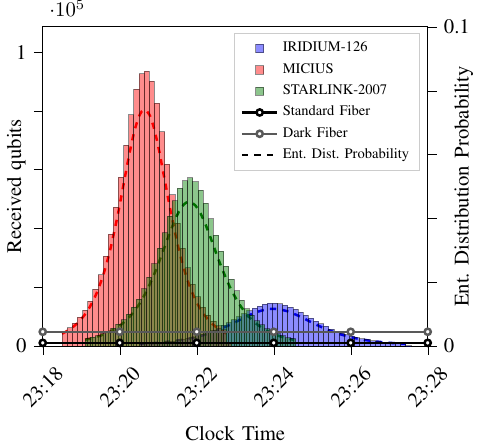}
    \caption{\small{Received qubits for various satellites overhead and fiber connections between Munich and Nuremberg with an average frame inter-arrival time of $20$ ms between frames.}}
    \label{fig:result1}
\end{figure}

In Fig.~\ref{fig:result1}, we present a comparison in terms of the number of received qubits between satellite and optical fiber link, binned in 8-second intervals with unlimited quantum storage. Firstly, Micius and Starlink-2007 offer a higher entanglement service during the time they are in visibility compared to Iridium-126. This is attributed to the higher altitude orbit of Iridium-126. Secondly, while both standard and dark fiber offer a continuous entanglement distribution, their performance are lower compared to satellite backbone.

\begin{figure}[t]
    \centering    
    \includegraphics[]{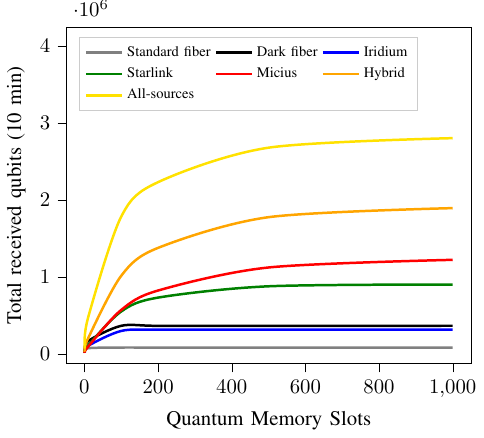}
    \caption{\small{Total received qubits in a service time window of $10$ minutes with varying memory.}}
    \label{fig:result2}
\end{figure}

In Fig.~\ref{fig:result2}, we compare the total amount of qubits received under finite memory sizes during a service time window of $10$ minutes. Specifically, Micius and Starlink-2007 offer a higher entanglement distribution rate compared to fiber backbone when $M \geq 20$. Iridium-126, instead, does not outperform dark-fiber. However, to provide a continuous and robust entanglement service, we envision that ground stations can choose, based on the current channel attenuation, from which quantum source they receive entangled photons, or, alternatively, simultaneously accept them from all the sources. Thus, to provide a continuous and robust entanglement service, a hybrid approach may be used in the near-term and an all-sources approach in the future. The near-term approach leverages both satellite and fiber backbones strategically, increasing the entanglement distribution rate based on real-time channel conditions and resources availability by considering the channel attenuation and satellite visibility to select the best entanglement source at each moment. Overall, both strategies enhance the overall reliability and performance of the entanglement service. Further, we see that under this configuration, at roughly 1,000 memory units, all trends converge implying additional memories will not improve the rate.

\begin{table*}
\centering
\resizebox{\linewidth}{!}{%
\input{plots/sat_table_data_1}
}
\caption{Satellite orbits data. The start and end dates denote the duration in which the satellites are above $20^\circ$ from both Munich and Nuremberg.}
\label{tab:sats}
\end{table*}

%% file: plots/sat_table_data_1.tex
\begin{tabular}{|c|c|c|c|c|c|c|} 
\hline
\multirow{2}{*}{Satellite Orbit} & \multirow{2}{*}{Start date, UTC} & \multirow{2}{*}{End date, UTC} & \multicolumn{2}{c|}{Max Elevation Angle ($\circ$)} & \multicolumn{2}{c|}{Minimum altitude (km)}  \\ 
\cline{4-7}
                                 &                                  &                                & Munich & Nuremberg                                  & Munich & Nuremberg                          \\ 
\hline
Micius                           & 26/02/2024, 23:18:30                & 26/02/2024, 23:22:46              & 83     & 75                                        & 474    & 486                                \\ 
\hline
Starlink-2007                   & 26/02/2024, 23:19:06                 & 26/02/2024, 23:24:32               & 88     & 75                                         & 551    & 567                               \\ 
\hline
Iridium-126                     & 26/02/2024, 23:20:30                & 26/02/2024, 23:27:26               & 76     & 74                                         & 804    & 809                                \\
\hline
\end{tabular}

%% file: Sections/06_challenges.tex
The design's practical application is to facilitate large-scale quantum networks, combining the advantages of scale-free networks at the metro scale and overcoming distance limitations through a quantum backbone network. Although the design enables this, it depends on technology that currently cannot facilitate the essential functions needed to fulfill the requirements. However, the long-term vision for quantum networks anticipates the incorporation of these capabilities over time. Several main hurdles must be overcome to achieve this vision.

The first requirement is an indexable quantum memory with extended storage lifetimes. Building such memories has posed a significant technological challenge, and current technology has not advanced sufficiently for commercial use. Yet, robust quantum memories are indispensable for large-scale quantum networks requiring quantum repeaters, serving as a key enabling technology for long-distance quantum communication.

Beyond robust storage, the memory must efficiently read and write quantum states and potentially convert an optical quantum state into another qubit type and back through transduction. Achieving this for quantum memory input-output demands technological advancements, particularly in the development of quantum transducers and quantum switches with low insertion loss~\cite{lauk2020perspectives}.

Another key requirement is the ability to perform  quantum teleportation reliably. Quantum teleportation requires a Bell-state measurement and with current linear optical technology, a success probability of roughly 57\% has recently been observed \cite{bayerbach2023bell}, but that means at best, with state-of-the-art technology 43\% of attempts fail. The optical technology for performing Bell-state measurements needs substantial advancement before it ceases to be a main hindrance.  

High-precision time synchronization is crucial among network edges, the ultimate consumers of distributed entanglement, with an accuracy requirement as stringent as 50 ps. The accuracy of time synchronization can be affected by signals traversing devices causing serial latency or jitter. Implementing highly accurate clock synchronization in constellations demands careful consideration due to these factors.

Finally, at the metro scale, using hybrid quantum dataframes requires the use of all-optical networks. Operating all-optical networks has proven in the past, using classical data payloads, to be challenging, and it becomes more difficult in the quantum regime \cite{diadamo2022packet}. Still, as quantum communication networks develop, the all-optical networks are seeing a higher level of attention and development.

 Overall, there are various technological milestones to reach before long-range, reliable, quantum networks can be deployed. Although these challenges exist, the quantum network community is optimistic there will be solutions in the future, and research effort in both academic and commercial is ongoing with strong efforts.

%% file: Sections/07_conclusion.tex
Overall, we propose a quantum backbone network design for hybrid quantum frame transmission to enable large-scale quantum communication networks. To achieve this, we introduce a network interface design that merges packetized quantum networks with an entanglement-based backbone network. Various link types were analyzed for the backbone network and a performance comparison of our design, in terms of the number of qubits received under different quantum memory sizes and fiber optical links, was carried out. The results show that satellite quantum backbone networks can offer an higher entanglement distribution service compared to the fiber one. However, since their service time strictly depends on their passage a time, an hybrid approach could lead to an optimal and robust quantum communication rate between subnetworks. As future work, we plan to investigate the potential of using \glspl{UAV} as part of the quantum backbone network architecture, exploring their feasibility and efficiency in enhancing the transmission rate.

%% file: ref.bbl

%% file: bios.tex
\begin{IEEEbiographynophoto}{Francesco Vista}
received the B. Sc and M. Sc degree in Computer Science Engineering from Politecnico di Bari, Italy, in 2020. Since 2020, he has been a PhD student at Politecnico di Bari and currently he is a quantum research intern at Cisco Quantum Lab.
\end{IEEEbiographynophoto}

\begin{IEEEbiographynophoto}{Daniel Holme} studied computer science at Nottingham Trent University until 1999, received an MBA from the Open University in 2017, and is  pursuing a M. Sc in Quantum Technology and Management at the University of Sussex. He is currently a Technology Lead for Global Industry Partners at Cisco Systems.
\end{IEEEbiographynophoto}

\begin{IEEEbiographynophoto}{Stephen DiAdamo}
received a Hon. B. Sc in computer science at the University of Toronto in 2014 and a M. Sc in mathematics in 2018 and Dr.-Ing in electrical engineering from the Technical University of Munich in 2023. Currently, he is a quantum research scientist at Cisco Quantum Lab.
\end{IEEEbiographynophoto}